\documentclass[11pt]{article}

\usepackage{amsmath}
\usepackage{amsfonts}
\usepackage{amssymb}
\usepackage{latexsym}
\usepackage{graphicx}
\usepackage{setspace}

\usepackage[american]{babel}

\usepackage{authblk}

\newtheorem{theorem}{Theorem}
\newtheorem{Remark}{Remark}
\newtheorem{corollary}{Corollary}

\newtheorem{definition}{Definition}
\newtheorem{example}{Example}

\begin{document}

\title{Cumulative Conditional Expectation Index}
\author[a]{M. Fern\'andez\thanks{Corresponding author.\\ E-mail addresses: mariela@ime.unicamp.br (M. Fern\'andez), veronica@ime.unicamp.br (V. A. Gonz\'alez-L\'opez). }}
\author[b]{V. A. Gonz\'alez-L\'opez}
\date{}
\affil[a]{University of Campinas, 
Brazil}
\affil[b]{Department of Statistics, University of Campinas, 
Brazil}
\maketitle



\abstract{In this paper we study the cumulative conditional expectation function (CCEF) in the copula context. It is shown how to compute CCEF in terms of the cumulative copula function, this natural representation allows to deduce some useful properties, for instance with applications to convex combination of copulas. We introduce approximations of CCEF based on Bernstein polynomial  copulas. 
We introduce estimators for CCEF, which were constructed through Bernstein polynomial estimators for copulas. The estimators are asymptotically normal and biased for CCEF. 

\section{Introduction}

In this paper we explore $\mathbb{E}[V\vert U \leq u]$ as a function of $u \in (0,1),$ we denote this quantity as {\it{cumulative conditional expectation function}}. One of the motivations of working with this quantity is its use for making decisions in real problems. 
Our target is to give the foundation for the construction of approximations and estimators for the cumulative conditional expected function. Also, we present in the first case the rate of approximation and in the second, the asymptotic distribution of the empirical process related to the estimator. \\

\section{Cumulative conditional expectation}\label{concept}

For some applied problems, it is helpful knowing the mean performance of one variable conditioned to another variable restricted to a region and not just to a single observation. In order to reach that knowledge we study the following measure. 
\begin{definition}Let $(U,V)$ be a random vector with associated 2-copula $C.$ Then the   cumulative conditional expectation function of $V$ on $U$ is given by $R_C(u)=\mathbb{E}[V \vert U \leq u], \,\,\, \forall u \in (0,1).$
\end{definition}
\begin{Remark}
\noindent  The cases $u=1$ and $u=0$ are excluded because they restrict themselves to the copula regression function. Indeed $\mathbb{E}[V \vert U \leq 1]=\mathbb{E}[V]=0.5$ is trivial and $\mathbb{E}[V \vert U \leq 0]=\mathbb{E}[V \vert U=0]$ has  been  studied in \cite{sungur2}. 
\end{Remark}
\noindent We note the previous concept was already explored in \cite{MFandVG-L2014}, but  that work focused on some bounds of the measure $R_C$. The following theorem gives a representation for $R_C(u)$ that let us to simplify its computation. 

\begin{theorem}\label{le:integral}
Let $(U,V)$ be a random vector with associated 2-copula $C.$ Then $R_C(u)=1- \frac{1}{u}\int_0^1C(u,v)dv,$ $\forall u \in (0,1).$
\end{theorem}

\noindent In order to simplify the notation, hereafter we denote the partial derivatives as done in \cite{Janssen2014}, this is $C^{(1)}(u,v)=\frac{\partial C(u,v)}{\partial u}$, $C^{(2)}(u,v)=\frac{\partial C(u,v)}{\partial v}$, $C^{(1,1)}(u,v)=\frac{\partial^2 C(u,v)}{\partial u^2}$ and $C^{(2,2)}(u,v)=\frac{\partial^2 C(u,v)}{\partial v^2}$. It can be seen that $\mathbb{E}[V\vert U \leq u ] $ is  the average of the copula regression function $\mathbb{E}[V \vert U=u]$.

\begin{corollary}
Under the hypotheses of theorem \ref{le:integral}, $R_{C}(u)= \frac{1}{u} \int_0^u \mathbb{E}[V \vert U=w] dw,\,\,\,\, \forall u \in (0,1).$
\end{corollary}
\noindent{ \it {Proof.~}}
By Theorem \ref{le:integral} we have $R_C(u)=1- \frac{1}{u}\int_0^1C(u,v)dv	$, then
$$
	R_C(u)= \frac{1}{u}\int_0^u dw -\frac{1}{u}\int_{0}^1 \int_0^u C^{(1)} (w,v)dw dv \\
			= \frac{1}{u}\int_0^u \Big(1- \int_0^1 C^{(1)}(w,v)dv\Big) dw.
$$
The last term is equal to $\frac{1}{u} \int_0^u \mathbb{E}[V \vert U =w] dw$ according to \cite{sungur2}.$\hfill \blacksquare$

By means of Theorem \ref{le:integral} we can compute the cumulative conditional expectation function for several cases, as it was explored in \cite{MFandVG-L2014}. 
The following result shows that the cumulative conditional expectation of a convex mixture of co\-pu\-las is the convex mixture of the cumulative conditional expectations. 

\begin{corollary}\label{cor:convex}Under the hypotheses of theorem \ref{le:integral}, let $C$ be the convex combination copula, $C(u,v)=\sum_{\gamma=1}^\Gamma p_{\gamma}C_{\gamma}(u,v)$ where $C_{\gamma}$ is a copula function, $0 \leq p_{\gamma} \leq 1$ for $1 \leq \gamma \leq \Gamma$ and $\sum_{\gamma=1}^{\Gamma} p_{\gamma}=1.$ Then, $R_C(u) = \sum_{\gamma=1}^{\Gamma} p_{\gamma} R_{C_{\gamma}}(u)$.
\end{corollary}


\noindent In the next result we study a specific class of copulas, the polynomial ones. Recalling that a copula has polynomial cross sections in $u$ if it can be written as $C(u,v)=\sum_{i=1}^k \alpha_i(v) u^i$ for each $u \in [0,1],$ to suitable functions $\alpha_i, i=1,\ldots, k.$ More details about these copulas can be found in \cite{nelsen} \S3.2. As an immediate consequence of Theorem \ref{le:integral} we have the following corollary.

\begin{corollary}
Under the hypotheses of theorem \ref{le:integral}, if $C(u,v)=\sum_{i=1}^k \alpha_i(v) u^i.$  Then, $R_C(u)$ is a polynomial of degree $k-1$ given by $R_C(u) = 1-\sum_{i=0}^{k-1} u^i \int_0^1 \alpha_{i+1}(v)dv.$
\end{corollary}

\begin{example}
Quadratic and cubic cross sections copulas.
\begin{enumerate}
\item  The Farlie-Gumbel-Morgenstern (FGM) family contains all the copulas with quadratic cross sections in both variables. $C(u,v)=uv +\theta uv(1-u)(1-v)$ with $\theta \in [-1,1],$ then $R_C(u)=\big(3+\theta (u-1)\big)/6$.
\item The Lin's iterated FGM family is defined by $C(u,v) = uv + \theta uv(1-u)(1-v) \big( 1+\varphi  (1-u)(1-v) \big)$ with $\theta$ and $\varphi$ real numbers satisfying $\theta \in [-1,1]$ and $-1-\theta \leq \theta(1+\varphi) \leq \big(3-\theta+(9-6\theta-3\theta^2)^{1/2}\big)/2.$ $C$ has cubic cross sections in both variables. For $u \in [0,1],$  $C(u,v)=\sum_{i=1}^3 \alpha_i(v) u^i$ where 
\begin{eqnarray*}
\alpha_1(v) &=& (1+\theta +\varphi \theta) v+ (-\theta -2\varphi  \theta) v^2 +\varphi  \theta v^3 \\
\alpha_2(v) &=& (-\theta -2\varphi  \theta)v + (\theta +4\varphi  \theta) v^2 - 2 \varphi  \theta v^3 \\
\alpha_3(v) &=& \varphi  \theta v -2 \varphi  \theta v^2 +\varphi  \theta v^3.
\end{eqnarray*}
Then, $R_C(u)=(\frac{1}{2} -\frac{\theta}{6} -\frac{\varphi  \theta}{12}) + (\frac{\theta}{6}+\frac{\theta \varphi }{6})u - \frac{\varphi \theta}{12} u^2$.
\end{enumerate}
\end{example}

\section{Bernstein Copulas}\label{sec:polynomial}

\begin{definition}\label{eq:BC}
Given a 2-copula $C$ and a grid of points $(u_k,\ v_l) \in [0,1]^2$ with $k,l=1, \dots, m,$ the Bernstein copula approximation of $C$ of order $m$ is given by $$B_mC(u,v)=\sum_{k=1}^m\sum_{l=1}^m C\Big(\frac{k}{m}, \frac{l}{m}\Big)P_{k,m}(u)P_{l,m}(v),\,\,\forall (u,v) \in[0,1]^2,$$ where $P_{j,m}(x)=\binom{m}{j} x^j(1-x)^{m-j},\,\,x \in [0,1].$ 
\end{definition}

The function $B_mC$ approximates the copula $C$ and is a copula itself. 
The usefulness of the next theorem is to  numerically simplify the computation of $R_C$, through solving polynomial integrals, when $C$ is a parametric copula described by a complex analytical expression.
\begin{definition}\label{eq:aproximaR} 
Consider a 2-copula $C,$ and a grid of points $(u_k,\ v_l) \in [0,1]^2$ with $k,l=1, \dots, m.$ Set the Bernstein approximation of $R_C$ of order $m$ as
\begin{equation*}\label{eq:aproximaR}
R_{B_mC}(u)=1-\frac{1}{(m+1)u}\sum_{k=1}^m\sum_{l=1}^m C\Big(\frac{k}{m}, \frac{l}{m}\Big)P_{k,m}(u),\,\,\,\forall u \in (0,1),
\end{equation*}
where $P_{k,m}(x)=\binom{m}{k} x^k(1-x)^{m-k},\,\,x \in [0,1].$
\end{definition}

\begin{theorem}\label{teo:converg}
Under the hypotheses of theorem \ref{le:integral}, $R_{B_mC}(u)$  converges to $R_C(u)$  pointwise, when $m \to \infty .$
\end{theorem}

\noindent For appropriate  copulas it is possible to give a rate of the approximation for $R_C$, and also we observe that this rate is improved when $u$ increases.

\begin{theorem}\label{teo:unif_converg}
Let $C$ be a continuous copula with first order partial derivatives being Lipschitz and $u_0 \in (0,1)$. For $u \in [u_0,1)$ it holds  $\big|R_{C}(u) - R_{B_mC}(u)\big|  \leq  \frac{7M}{12u_0m}$ for a constant $M$. Then,  $R_{B_mC}$ converges uniformly to $R_C$ in $[u_0,1)$, when $m \to \infty .$
\end{theorem}

\section{Estimation}\label{sec:estimation}
Consider a bivariate random sample $\{ (U_j,V_j)\}_{j=1}^n$ of the vector $(U,V)$ with associated unknown 2-copula $C.$ Let be $R_{U_j}=$ rank of $U_j$ in $\left\{ U_1,\ldots, U_n\right\}$ and $R_{V_j}=$ rank of $V_j$ in $\left\{ V_1,\ldots, V_n\right\}$.
Denote the empirical copula as $$C_n(u,v)=\frac{1}{n} \sum_{j=1}^n  I\Big( \frac{R_{U_j}}{n} \leq u\Big) I\Big(\frac{R_{V_j}}{n} \leq v\Big),\,\,\,\forall u, v \in [0,1].$$  By using the Bernstein estimator of the copula function $C$ proposed in \cite{sancetta} we obtain the estimator of the cumulative conditional expectation function.
\begin{definition}\label{eq:estimador}
Given a 2-copula $C,$ the estimator of $R_C(u)$ is
\begin{equation*}\label{eq:estimador}
\hat{R}_{B_mC_n}(u)=1-\frac{1}{(m+1)u}\sum_{k=0}^{m}\sum_{l=0}^{m}C_n\Big(\frac{k}{m},\frac{l}{m}\Big)P_{k,m}(u), \,\,\, \forall u \in (0,1),
\end{equation*}
 where $C_n$ is the empirical copula. 
\end{definition}

In order to explore the asymptotic behavior of the estimator $\hat{R}_{B_mC_n}(u)$ we make use of some known results about the Bernstein estimator. In \cite{Janssen2012} (Theorem 2 and Remark 3) it is proved that for a copula $C$ with  bounded third order partial derivatives on $(0,1)^2,$ if $\sqrt{n}/m \rightarrow d$, $0 \leq d < \infty,$ when $n \to \infty ,$ then the process  $\sqrt{n}\Big( B_mC_n(u,v) - C(u,v) \Big)\leadsto {\mathcal{G}}_{C}(u,v)$ in the space $l^{\infty}\big((0,1)^2\big)$ of bounded functions, where $\leadsto$ denotes weak convergence. In addition, the limiting process ${\mathcal{G}}_{C}(u,v)$ is a tight Gaussian process with mean function  $db(u,v)$ where
\begin{equation}\label{funcionb}
b(u,v)=\frac{1}{2}\Big(u(1-u)C^{(1,1)}(u,v)+v(1-v) C^{(2,2)}(u,v)\Big)
\end{equation}
and covariance function given by $E[h(u,v)h(u^{\prime},v^{\prime}) ]$ for $0<u,\ u^{\prime},\ v,\ v^{\prime}<1$ with
\begin{equation}\label{funcionh}
h(u,v)= I(U \leq u, V \leq v)-C(u,v)- C^{(1)}(u,v)\big(I(U \leq u)-u\big) -  C^{(2)}(u,v)\big(I(V \leq v)-v\big).
\end{equation}

This time we consider the process $\frac{1}{u}\int_{0}^{1}{\mathcal{G}}_{C}(u,v)dv, $ with $u \in (0,1),$ which is the process that defines the asymptotic behavior of the proposed estimator of $R_C.$

\begin{theorem}\label{teo:asypmtotic}
Let $C$ be a 2-copula  with  bounded third order partial derivatives on $(0,1)^2$. If $n$ and $m$ are positive integers such that $\sqrt{n}/m \rightarrow d$ with $0 \leq d < \infty$ when $n, m \to \infty,$ then for $u \in (0,1)$,
$$
\sqrt{n}\Big( \hat{R}_{B_mC_n}(u)-R_{C}(u)\Big) \leadsto -\frac{1}{u}\int_{0}^{1}{\mathcal{G}}_{C}(u,v)dv.
$$ 
$-\frac{1}{u}\int_{0}^{1}{\mathcal{G}}_{C}(u,v)dv$ is a Gaussian process with mean function $d(\frac{1}{2}-R_C(u))+\frac{d(u-1)}{2}\int_{0}^{1}C^{(1,1)}(u,v)dv$ and
variance function
\begin{eqnarray*}
\hspace{-0.5cm}Var\Big[-\frac{1}{u}\int_{0}^{1}{\mathcal{G}}_{C}(u,v)dv\Big]&=&-4R_C^2(u)+\big(4-\frac{1}{u}\big)R_C(u)-4\big(2-\frac{1}{u}\big)R_C(u)\mathcal{H}_1(u,u)\\
&&\hspace{-1.5cm}+2\big(3-\frac{2}{u}\big)\mathcal{H}_1(u,u)-4\big(1-\frac{1}{u}\big)\mathcal{H}_1^2(u,u)-2+\frac{1}{u}+\frac{1}{u^2}\int_{0}^{1}\big(C^{(2)}(u,v)\big)^2vdv. 
\end{eqnarray*}
where $2 \mathcal{H}_1(u,u)=\int_0^1C^{(1)}(u,v)dv.$
\end{theorem}

\begin{Remark}\label{remark:covar}
The covariance function of the process $-\frac{1}{u}\int_{0}^{1}{\mathcal{G}}_{C}(u,v)dv$  is 
\begin{eqnarray*}
\hspace{-1cm}Cov\Big[-\frac{1}{u}\int_{0}^{1}{\mathcal{G}}_{C}(u,v)dv,-\frac{1}{u'}\int_{0}^{1}{\mathcal{G}}_{C}(u',v')dv'\Big]=\\
&&\hspace{-7cm}=\frac{1}{uu'}\int_{0}^{1}\int_{0}^{1}C( u\wedge u', v\wedge v')dv'dv+\big(1-R_C(u)\big)\big(R_C(u')-1\big)+\mathcal{H}_3(u,u') +\mathcal{H}_3(u',u)\\
&&\hspace{-7cm}+\big( \frac{1}{u\vee u'}-1\big)\int_0^1C^{(1)}(u',v)dv\int_0^1C^{(1)}(u,v)dv -\mathcal{H}_2(u,u')-\mathcal{H}_2(u',u)\\
&&\hspace{-7cm}+\frac{1}{uu'}\int_{0}^{1}\int_{0}^{1}C^{(2)}(u',v')C^{(2)}(u,v)(v\wedge v')dvdv'-R_C(u)R_C(u')+\mathcal{H}_1(u,u')+\mathcal{H}_1(u',u).
\end{eqnarray*}
with
\begin{eqnarray*}
\mathcal{H}_1(w_1,w_2) &=& \frac{1}{w_1w_2}\int_0^1C^{(1)}(w_1,v)dv\int_{0}^{1}C( w_1,v)C^{(2)}(w_2,v)dv\\
\mathcal{H}_2(w_1,w_2) &=& \frac{1}{w_1w_2}\int_{0}^{1}\int_{0}^{1}C^{(2)}(w_2,v)C(w_1,v\wedge v')dvdv'-\big(1-R_C(w_1)\big)R_C(w_2)\\
\mathcal{H}_3(w_1,w_2)&=&\Big(\frac{1}{w_1\vee w_2}\big(R_C(w_1\wedge w_2)-1\big)+1-2R_C( w_1)\Big)\int_0^1C^{(1)}(w_2,v)dv.
\end{eqnarray*}
for $w_1\vee w_2 = \max\{w_1,w_2\}$ and $w_1\wedge w_2 = \min\{w_1,w_2\}$. 
\end{Remark}


\end{document}